\documentclass[]{aipproc}
\layoutstyle{6x9}
\usepackage{amsmath,amssymb}
\usepackage{epsfig}

\def\tsig{{\ensuremath\mathbf{\sigma}}}
\def\tchi{{\ensuremath\mathbf{\chi}}}
\def\ttau{{\ensuremath\mathbf{\tau}}}

\begin{document}
\title{Transport of Strongly Correlated Electrons}
\author{P. Prelov\v sek}{address=
{Faculty of Mathematics and Physics, University of
Ljubljana, and J. Stefan Institute, 100 Ljubljana, Slovenia }}
\author{X. Zotos}{address=
{Institut Romand de Recherche Num\'erique en Physique des
Mat\'eriaux (EPFL-PPH), CH-1015 Lausanne, Switzerland }}
%\date{\today}
\begin{abstract}
Lectures deal with the theory of electronic transport, in particular
with the electrical conductivity, in systems dominated by strong
electron-electron repulsion. The concept of charge stiffness is
introduced to distinguish conductors and insulators at $T=0$, but as
well usual resistors, possible ideal conductors and ideal insulators
at finite temperature. It is shown that the latter singular
transport appears in many integrable systems of interacting fermions,
the evidence coming from the relation with level dynamics, from the
existence of conserved quantities as well as from numerical studies
and exact results. Then, exact duagonalization approaches for the
calculation of static and dynamical quantities in small correlated
systems are described, with the emphasis on the finite-temperature
Lanczos method applicable to transport quantities. Finally, anomalous
dynamical conductivity within the planar single-band model is
discussed in relation with experiments on cuprates.

\end{abstract}
\maketitle

\section{Introduction}

In the theory of strongly correlated electrons there are at present
still numerous open problems, both regarding the basic understanding
of phenomena and even more of the methods of calculation and sensible
approximations. Among such theoretical challenges are also electronic
transport quantities, such as electrical conductivity, the electronic
heat conductivity, the thermopower and the Hall constant. In usual
metals and semiconductors the transport theory, based on the Boltzmann
semiclassical approach as well as the quantum calculation of relevant
scattering mechanisms, involving impurities and electron-phonon
coupling, date back to the beginning of then solid state theory. The
extention of these methods to strong disorder and strong
electron-phonon coupling, introducing the phenomena of localization,
polaronic transport etc., using the framework of general linear
response theory have been in the focus of theoreticians for several
decades.  Within the same treatment also the effects of the weak
electron-electron coupling has been understood, as summarized within
the Landau Fermi liquid concept for electron transport. On the other
hand, the strong electron-electron Coulomb repulsion, being the
signature of strongly correlated systems, raises new questions
of appropriate theoretical techiques and moreover of the basic
understanding of the scattering mechanism and its effects.

As a prominent example of the transport quantity we will mainly focus
on {\it the dynamical (optical) electrical conductivity}
$\sigma(\omega)$, whereby the easiest accesible experimental
quantity is the d.c. resistivity $\rho=1/\sigma(0)$. Within the
usual Botzmann theory for a weak electron scattering the relaxation-time
approximation represents well the low-frequancy behaviour,
\begin{equation}
\sigma(\omega)=\sigma_0/(1+i\omega \tau),
\end{equation}
where the relaxation time $\tau$ depends on the particular scattering
mechanism and is in general temperature dependent. In the following we
will consider only homogeneous systems without any disorder, so the
relevant processes in the solid state are electron-phonon scattering
and the electron-electron (Coulomb) repulsion. When the latter becomes
strong it is expected to dominate also the transport quantities. In
this case there appear evidently fundamentally new questions:

a) The repulsion can transform a metal (conductor) into an
(Mott-Hubbard) insulator at $T=0$ which radically changes transport
quantities, both in the ground state as well as at finite temperatures
$T>0$.

b) Even in the metallic phase it is not evident which is the relevant
scattering process determining $\tau(T)$ and $\rho(T)$. In the absense
of disorder and neglecting the electron-phonon coupling the standard
theory of purely electron-electron scattering would state that one
needs Umklapp scattering processes to yield finite $\tau$.  That is,
the relevant electron Hamiltonian includes the kinetic energy, the
lattice periodic potential $V$ and the electron-electron interaction
$H_{int}$, 
\begin{equation}
H= H_{kin} + V +H_{int}.
\end{equation}
Then, in general the electronic current density $j$ is not conserved
due to Umklapp processes in the electron-electron scattering (where
the sum ef electron momenta is nonzero, equal to nonzero reciprocal
wavevector), i.e. $[H,j]\neq 0$, leading to the current relaxation and
dissipation. The interplay of $V$ and $H_{int}$ becomes however quite
involved in the case of a strong electron repulsion. This becomes
quite clear in examples of integrable tight binding models of
interacting fermions, which have anomalous (diverging) transport
quantities.

c) Experiments on cuprates, manganites and other novel materials -
strange metals - with correlated electrons question the validity of
the concept of the current relaxation rate $1/\tau$. For example, in
cuprates only strongly frequency (and $T$) dependent $\tau(\omega,T)$
can account for experiments in the normal state, leading to the
concept of the marginal Fermi liquid.

In these lectures we will consider only the physics of strongly
correlated electrons. The standard framework for theoretical
investigations in this case are microscopic tight-binding models for
electrons (in general with spin), including the hopping term $T$
(representing the effect of $V$ and $H_{kin}$ and the
electron-electron interaction $H_{int}$
\begin{equation}
\hat H=-\sum_{i,j,s} t_{ij}( c^{\dagger}_{js} c_{is} +H.c.) +H_{int}=
T+H_{int}, \label{mod}
\end{equation}
where $c_{is} ( c^{\dagger}_{is})$ are annihilation (creation)
operators for fermions on sites $i$ on a D-dimensional lattice and in
most cases $t_{ij}=t$ is the hopping matrix element between nearest
neighbor (n.n.) sites only. $H_{int}$ is a local (onsite or including
only few neighboring sites) interaction defined for specific cases
lateron.

\section{Linear response theory}

The linear response theory for the electrical a.c. conductivity
$\sigma(\omega)$ has been formulated by Kubo and is standard for
dissipative systems \cite{maha}. Nevetheless, in correlated systems at
$T=0$ as well as at $T>0$ in particular cases (of integrable models)
one has to take in general into account besides the regular part
$\sigma_{reg}(\omega)$ also the singular contribution to the charge
dynamical response, i.e.
\begin{equation}
\sigma(\omega)= 2\pi e_0^2 D \delta(\omega) +\sigma_{reg}(\omega),
\label{ec3}
\end{equation}
where $D$ represents the charge stiffness. The analysis in the
presence of finite $D>0$ as well as the derivation of the proper
optical sum rule within the models of correlated elctrons (\ref{mod})
requires more care and goes beyond the standard linear response
formulations, hence we present it shortly below.

We follow the approach by Kohn \cite{kohn} introducing a (fictitious)
flux $\phi$ through a torus representing the lattice with periodic
boundary conditions (p.b.c.). Such a flux induces a vector potential
${\bf A}$, being equal on all lattice sites.  In lattice
(tight-binding) models with a discrete basis for electron
wavefunctions the vector potential ${\bf A}$ can be introduced via a
gauge transformation (Peierls construction)
\begin{equation}
c^\dagger_{js}\to c^\dagger_{js} {\rm exp}(-ie_0{\bf A}\cdot{\bf R}_j),
\end{equation}
which effectively modifies hopping matrix elements $t_{ij}$.  Taking
${\bf A}$ as small and assuming that magnetic field does not modify
within the lowest order the (local) interaction term $H_{int}$ one can
express the modified tight-binding Hamiltonian (\ref{mod})
\begin{eqnarray}
H({\bf A})&=&-\sum_{i,j,s}
t_{ij}e^{-ie_0{\bf A}\cdot{\bf R}_{ij}}c^\dagger_{js}c_{is}+H_{int} 
\approx \nonumber \\
&\approx&H(0)+e_0{\bf A}\cdot{\bf j}+\frac{e_0^2}{2}
{\bf A}\cdot \ttau {\bf A} = H(0) + H' , \label{ec5}
\end{eqnarray}
where ${\bf R}_{ij}={\bf R}_j-{\bf R}_i$ and ${\bf j}$ and $\ttau$ are
the current and the stress tensor operators, respectively, given by
\begin{eqnarray}
{\bf j}&=&i\sum_{i,j,s}
t_{ij}{\bf R}_{ij} c^\dagger_{js}c_{is}, \nonumber \\
 \ttau&=&\sum_{i,j,s}
 t_{ij}{\bf R}_{ij}\otimes{\bf R}_{ij} c^\dagger_{js}c_{is}.
\label{ec6}
\end{eqnarray}
Note that in usual n.n. tight-binding models $\ttau$ is directly
related to the kinetic energy operator, $\ttau_{\alpha\alpha} =
(H_{kin})_{\alpha\alpha}$.

The electrical current ${\bf j}_e$ is from the equation (\ref{ec5})
expressed as a sum of the particle-current and the diamagnetic
contribution,
\begin{equation}
{\bf j}_e=-\partial H/\partial{\bf A}=-e_0{\bf j}-e_0^2\ttau{\bf A}.
\label{ec7}
\end{equation}
The above analysis applies also to an oscillating ${\bf A}(t) = {\bf
A}(\omega){\rm exp}(-i\omega^+ t)$ with $\omega^+=\omega+i\delta$ with
$\delta \to 0$. This induces an electric field in the system ${\bf
E}(t)= - \partial{\bf A}(t)/\partial t$.  We are interested in the
response of $\langle {\bf j}_e\rangle(\omega)$. Evaluating
$\langle{\bf j}\rangle$ as a linear response \cite{maha} to the
perturbation $H'$ in Eq.(\ref{ec5}), and with ${\bf A}(\omega)= {\bf
E}(\omega)/i\omega^+$, we arrive at the optical conductivity,
\begin{eqnarray}
\tilde {\tsig}(\omega)&=&\frac{i e_0^2}{\omega^+ N}
(\langle \ttau\rangle - \tchi(\omega)),\nonumber \\
\tchi(\omega)&=& i\int_0^{\infty} dt e^{i\omega^+t}\langle[{\bf j}(t), 
{\bf j}(0)]\rangle. \label{ec8}
\end{eqnarray}
Complex $\tilde {\tsig}(\omega)= {\tsig}(\omega)+i\tilde
{\tsig}^{\prime\prime}(\omega)$ satisfies the Kramers-Kronig
relation. Since $\tchi(\omega \to \infty) \to 0$, we get from the
equation (\ref{ec8}) a condition for $\tilde {\tsig}^{\prime\prime}
(\omega\to \infty)$,
\begin{equation}
\int_{-\infty}^\infty \tsig(\omega)d\omega=
\frac{\pi e_0^2}{N}\langle \ttau \rangle,  \label{ec9}
\end{equation}
which corresponds to the optical sum rule. It reduces to the well
known sum rule for continuum electronic systems, as well as for
n.n. hopping models where $\langle \tau_{\alpha\alpha} \rangle=
-\langle H_{kin}\rangle/d$ \cite{mald}.  We can now make contact
with the definition (\ref{ec3}). From the expression (\ref{ec8}) it
follows
\begin{equation}
D_{\alpha\alpha} =\frac{1}{2e_0^2}\lim_{\omega\to 0}\omega \tilde
\sigma_{\alpha\alpha}^{\prime\prime}(\omega)=
\frac{1}{2N}[\langle \tau_{\alpha\alpha}
\rangle-\chi_{\alpha\alpha}'(0)], \label{stif}
\end{equation}
and
\begin{eqnarray}
\sigma_{reg}(\omega)&=&\frac{e_0^2}{N\omega}\chi''_{\alpha\alpha}(\omega)=
e_0^2 \frac{1-{\rm e}^{-\beta\omega}}{\omega} C_{\alpha\alpha}(\omega),
\nonumber \\ C_{\alpha\alpha}(\omega)&=&\frac{1}{N}{\rm
Re}\int_0^\infty dt {\rm e}^{i\omega t} \langle j_\alpha(t) j_\alpha
(0)\rangle. \label{reg}
\end{eqnarray}   

In any finite system one can write $C_{\alpha\alpha}(\omega)$ and
$D_{\alpha\alpha}$ in terms of exact eigenstates $|n\rangle$ with
corresponding energies $\epsilon_n$,
\begin{eqnarray}
C_{\alpha\alpha}(\omega)&=& \frac{1}{N} \sum_{n\ne m} p_n |\langle
m|j_{\alpha}|n \rangle|^2 \delta(\omega - \epsilon_m+ \epsilon_n),
\nonumber \\
D_{\alpha\alpha}(\omega)&=& \frac{1}{N} \bigl[ \langle
\tau_{\alpha\alpha}\rangle - \sum_{n\ne m} p_n \frac{ |\langle
m|j_{\alpha}|n \rangle|^2} {\epsilon_m- \epsilon_n} \bigr],
\label{ec2}
\end{eqnarray}
where $p_n={\rm e}^{-\beta \epsilon_n}/Z$ is the Boltzmann weight of
each level and $Z=\sum_n{\rm e}^{-\beta \epsilon_n}$ is the
thermodynamic sum.

\subsection{Conductors and insulators}

\subsubsection{Ground state}

At $T=0$ the charge stiffness
$D^0_{\alpha\alpha}=D_{\alpha\alpha}(T=0)$ is the central quantity
determining the charge transport. As already formulated by Kohn
\cite{kohn} $D^0$ can be expressed directly as a sensitivity of the
ground state to the applied flux $\phi_\alpha=eA_\alpha$,
\begin{equation}
D^0_{\alpha\alpha}=\frac{1}{2N}\frac{\partial^2 E_0}{\partial
\phi_\alpha^2} \bigl|_{\phi_\alpha=0}. \label{stif0}
\end{equation}
Since at $T=0$ there cannot be any dissipation and one expects
$\sigma_{reg}(\omega \to 0)=0$, we have to deal with two fundamentally
different possibilities with respect to $D^0$ (for simplicity we
consider an isotropic tensor ${\bf D^0}$):

a) $D^0>0$ is characteristic for a {\it conductor} or {\it metal},

b) $D^0=0$ applies to an {\it insulator}, which can have the origin in
the filled band of electrons (usual band insulator) or for non-filled
band in the Mott-Hubbard mechanism due to strong electron repulsion,
i.e. electron correlations. The latter situation is clearly of
interest here. Note that the same criterion of the sensitivity to flux
has been be applied to disordered systems, relevant to the theory of
the electron localization.

The theory of the metal-insulator transition solely due to the Coulomb
repulsion (Mott transition) has been intensively investigated in last
decades and is one of better understood parts of the physics of
strongly correlated electrons. The emphasis has been on analytical and
numerical studies of particular models of correlated electrons.  The
prototype and the most investigated model in this context is the
Hubbard model with the onsite repulsion,
\begin{equation}
H_{int}=U \sum_i n_{i\uparrow} n_{i\downarrow}. \label{hub}
\end{equation}

Let us consider as an example the Hubbard model on a one dimensional
(1D) chain with $L$ sites. We first derive the result for free fermions
at $U=0$. The eigenvalues in the presence of a flux $\phi$
become $\epsilon_k=-2t\cos(k+\phi)$ giving:
\begin{equation}
E_0= -4 t\sum_{|k|<k_F} \cos(k+\phi),~~~k=2\pi~{\rm integer}/L.
\end{equation}
Following Eq.(\ref{stif0}) we get
\begin{equation}
D^0= - \frac{1}{L}\langle 0|T|0\rangle=\frac{2t}{\pi}\sin k_F=
\frac{2t}{\pi}\sin \frac{\pi n}{2},
\end{equation}
where $n$ is the density of fermions ($n=2 k_F/\pi$). We notice that
$D$ vanishes for an empty band $n=0$, i.e. $D^0\simeq tn$ for
$n\rightarrow 0$, and for a filled band $n=2$ where $D^0\simeq t(2-n)$
for $n\rightarrow 2$.  $D^0$ is maximum at half filling where
$D^0(n=1)=2t/\pi$.

Another simple limit is $U=\infty$. Since in this case the double
occupation of sites is forbidden due to Eq.(\ref{hub}),
i.e. $n_{i\uparrow} n_{i\downarrow}=0$, fermions behave effectively
as spinless and the result is
\begin{equation}
D^0=\frac{t}{\pi}|\sin (\pi n)|,
\end{equation}
and in contrast to noninteracting case $D^0$ vanishes at half filling,
i.e. $D^0\simeq t|1-n|$ for $n\rightarrow 1$.

Analytical results in 1D indicate that $D^0=0$ at half filling
persists in the Hubbard model at all $U>0$, whereby the dependence
$D(n)$ is between both limits $U=0$ and $U=\infty$. The insulating
state at half filling is a generic feature also for a wider class of
other 1D models, as the $t$-$V$ model (equivalent to the anisotropic
Heisenberg model), the $t$-$J$ model etc.

The metal - insulator transition in higher dimensions is more
difficult and open subject and there are no exact analytical results
\cite{imad}. Still it is established that in the 2D Hubbard model
there is an insulator at half filling for arbitrary $U$ due to the
nesting of the Fermi surface. The insulating character is explicitly
evident also within the 2D $t$-$J$ model, which represents the large
$U$ limit of the Hubbard model. The metal - insulator transition has
been also extensively studied in the limit of infinite dimensions
\cite{geor}.  One would like to know the behavior of weakly doped
Mott-Hubbard insulator at $n<1$. Numerical results for $D$ on Hubbard
model as well as the $t$-$J$ model in 2D seem to indicate linear
variation $D \propto (1-n)^\alpha$ with $\alpha=1$ \cite{dago}, but
there are also arguments for $\alpha>1$ \cite{imad}.
 
\subsubsection{Finite Temperature}

At $T>0$ there are in principle more possibilities, since we have to
deal with $\tilde \sigma_0= \sigma_{reg}(\omega \to 0)\geq 0$ as well
as with $D(T) \geq 0$ \cite{zp}:

a) {\it Normal conductor} or {\it resistor} has a nonsingular response,
i.e. $D=0$, and a finite conductivity (resistivity) $\tilde \sigma_0>0$.

b) $D(T)>0$ would be a signature of an {\it ideal conductor}, which in
analogy to $T=0$ does not show any Ohm's energy dissipation, but rather a
reactive response (acceleration) of the charge in an applied external
electric field. In this case $\tilde \sigma_0$ is irrelevant since
a singular $D>0$ term dominates at low $\omega$. We will see that such
a case is realized in a number of nontrivial integrable many-fermion
models.

c) The situation with $D(T)=0$ and $\tilde \sigma_0=0$ would
correspond to an {\it ideal insulator}, since there will not be no
conduction in spite of $T>0$. This is clearly a very unusual
possibility, for which so far at least indications exist in integrable
systems \cite{zp}.

\section{Transport and Integrability}

It is plausible that integrable systems can have anomalous transport
properties. In classical systems this observation is well established.
On the other hand, the proposition that integrable quantum many-body
systems show dissipationless transport at finite temperatures is
rather recent \cite{czp}. The key to this idea comes from a study of
the Drude weight at finite temperatures $D(T>0)$. Since all integrable
many-body fermionic models are in 1D, we will further restrict
ourselves to 1D, although some statements are more general.

\subsection{Transport and level dynamics} 

It is convenient to discuss $D(T)$ with a generalization of the
original Kohn's approach \cite{kohn} for zero temperature, by relating
$D(T)$ to the thermal average of curvatures $D_n$ of energy levels
$\epsilon_n$ subject to the fictitious flux $\phi=eA$. Since the flux
induces the perturbation $H'=-\phi j$, we can expand the eigenenergies
$\epsilon_n(\phi)$ up to the second order in $\phi $,
\begin{equation}
\epsilon_n(\phi)=\epsilon_n(0)-\phi \langle n|j|n \rangle +
\frac{1}{2} \langle n|j|n \rangle + \phi^2  \sum_{n\ne m} \frac{ |\langle
m|j|n \rangle|^2} {\epsilon_m- \epsilon_n}. \label{eps}
\end{equation}
Following Eq.(\ref{ec2})) we can then write $D(T)$ as
\begin{equation}
D = \frac{1}{L} \sum_n p_n D_n= \frac{1}{2L} \sum_n p_n \frac
{\partial ^2 \epsilon_n (\phi) }{\partial \phi^2}|_{\phi \rightarrow
0}~, \label{eq5}
\end{equation}
i.e. in analogy to $T=0$ case $D$ can be expressed in terms of
curvatures of levels, at $T>0$ weighted with their Boltzmann factors.

Evaluating the second derivative of the free energy $F$ as a 
function of flux $\phi$, we also find
\begin{equation}
\frac{\partial^2 F}{\partial\phi^2}=
2D-\beta\sum_n p_n\left(\frac{\partial\epsilon_n}
{\partial\phi}\right)^2+\left(\sum_n p_n
\frac {\partial\epsilon_n}{\partial \phi}\right)^2. \label{eq6}
\end{equation}
In this expression the third term vanishes by symmetry and in a system
as ours without persistent currents, ${\partial^2 F/\partial\phi^2}$
has to vanish in the thermodynamic limit.  Therefore in the $L\to
\infty $ limit the charge stiffness can be alternatively represented as
\begin{equation}
D = \frac{\beta}{2L}\sum_n p_n \left(\frac{\partial\epsilon_n}{\partial\phi}
\right)^2. \label{eq7}
\end{equation}

On the other hand, we know from rather recent studies that the
integrability of a system is reflected in the distribution of level
spacings $P(s)$ \cite{ls}.  For {\it integrable systems} $P(s)$
follows the {\it Poisson distribution} $P(s)\propto exp(-s)$, while in
{\it non-integrable systems the Gaussian orthogonal ensemble} (GOE)
one given by the Wigner surmise, $P(s)\propto s~ exp(-s^2)$. The
difference is due to the existence of a macroscopic number of
conservation laws in integrable systems that, allowing level crossing,
supresses level repulsion.  As for the behavior of $D$, we can now
argue that in non-integrable systems, due to the level repulsion, the
levels move as a function of $\phi$ in an energy spacing of the order
of the inverse of the many - body density of states. The latter
becomes exponentially small with increasing $L$, leading to an
exponential suppression of slopes $\partial\epsilon_n/\partial\phi$ as
well as $D(T)$.  On the other hand, in the integrable system, the
slopes due to the absence of level repulsion can take values of the
order of one and $D(T)$ can remain finite even in the thermodynamic
limit.

In non-integrable systems the assumption of the random matrix theory
has also other consequensces. In particular, there exists a relation
between off-diagonal and diagonal elements \cite{wilk}
\begin{equation}
\langle\mid j_{n m}\mid^2\rangle_{\epsilon_n\neq\epsilon_m}=
\langle\left(j_{n n}-\bar{v}(E)\right)^2\rangle=\sigma^2(E).
\label{eq8}
\end{equation}
Here $j_{n m}=\langle n\mid j\mid m\rangle$, $\bar{v}=\langle
j_{n n}\rangle$ and the outer brackets denote an average of matrix
elements locally in the spectrum.  This leads for a normal resistor
with $D(T>0)=0$ to the relation for the d.c. conductivity,
\begin{equation}
\sigma_0=\beta \pi e^2 \sum_{n\ne m} p_n \mid j_{n m}\mid^2
\delta(\epsilon_n - \epsilon_m) 
= \frac{\beta\pi e^2}{ 2}\int_{-\infty}^{\infty} \frac{\exp(-\beta E)}
{Z} \rho^2(E)\sigma^2(E)dE. \label{eq9}
\end{equation}
$\sigma_0$ can be thus expressed in terms of $\sigma(\epsilon)$, which
measures the sensitivity to flux (velocity of level dynamics). Such an
approach has been used already in the theory of conduction in
disordered systems.

\subsection{Models}

\subsubsection{Particle in a fermionic bath}

The first model we study is a toy model of dissipation. The Hamiltonian
describing a single (tagged) particle interacting with a bath of spinless 
fermions is
\begin{equation}
H=-t_h\sum_i (e^{i\phi} d^{\dagger}_{i+1} d_i +H.c.)
-t\sum_i (c^{\dagger}_{i+1} c_i + H.c. ) +
U\sum_i d^{\dagger}_i d_i c^{\dagger}_i c_i ~,\label{fmb}
\end{equation}
where $c_i ( c^{\dagger}_i )$ are operators for spinless fermions
(without charge, i.e. not sensitive to the flux) and $d_i (
d^{\dagger}_i )$ for the tagged particle on a chain with p.b.c. The
interaction comes only through the on-site repulsion $U>0$. The
current $j$ refers to the tagged particle only, 
\begin{equation}
j=-it_h\sum_i d^{\dagger}_{i+1} d_i + H.c.
\end{equation}
and is not conserved for $U>0$, i.e. $[j,H]\neq 0$. In this model the
conductivity $\sigma(\omega)$ corresponds to the mobility
$\mu(\omega)$ of the tagged particle. The model (\ref{fmb}) is
integrable by the Bethe ansatz method in the case of equal masses,
i.e.  $t_h = t$ (where it is equivalent to the problem of a Hubbard
chain in a nearly polarized state $S^z=S_{max}^z-1$) and
non-integrable for unequal masses $t_h\neq t$.

\begin{figure}
\centering
%\resizebox{.7\columnwidth}{!b}
\epsfig{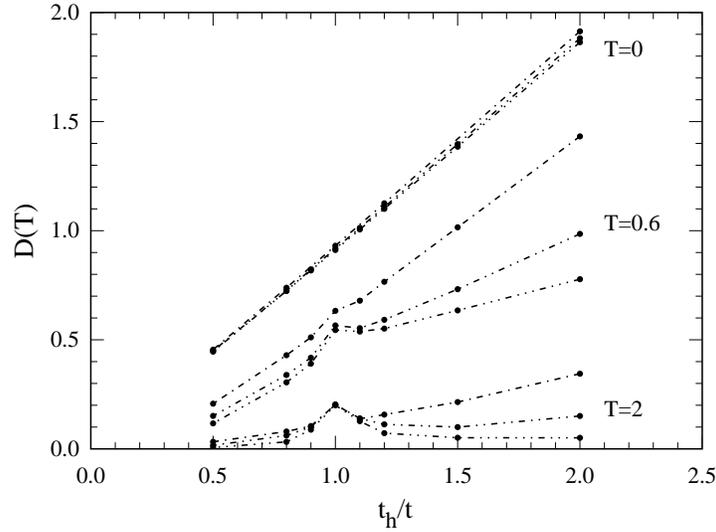}
%{\includegraphics{figx1.ps} }
%\hspace {0cm} {\epsfxsize 10cm \epsfysize 8cm \epsffile{figx1.ps}}
\caption{
$D(T)/D_0$ as a function of $t_h/t$ for $U/t=2$,
different temperatures $T$ and system sizes $L$:
$( -~\cdot~- )~ L=6$,
$( -~\cdot~\cdot~- )~ L=10$,
$( -~\cdot~\cdot~\cdot~- )~ L=14$.}
\label{Fig. 1}
\end{figure}

In Fig.~1 we present $D(T)$ as a function of $t_h/t$ \cite{czp}, for
different $T$ and $L$ but for fixed $U/t=2$ at half-filled band of
spinless electrons $N_e=L/2$. Results clearly reveal:

a) a nonmonotonic behavior of $D(T)$ as we sweep through the
integrable $t_h=t$ point for $T>0$,

b) a very weak dependence of $D(T)$ on the system size $L$ for the
integrable case $t_h=t$, and on contrary a rather strong one for the
non-integrable ones.

To obtain an impression of the overall behavior of $D(T)$ as a
function of $T$, we present in Fig.~2 $D(T)/D_0$ for $t_h/t=1.0, 0.5$
and different size systems.

\begin{figure}
\centering
\epsfig{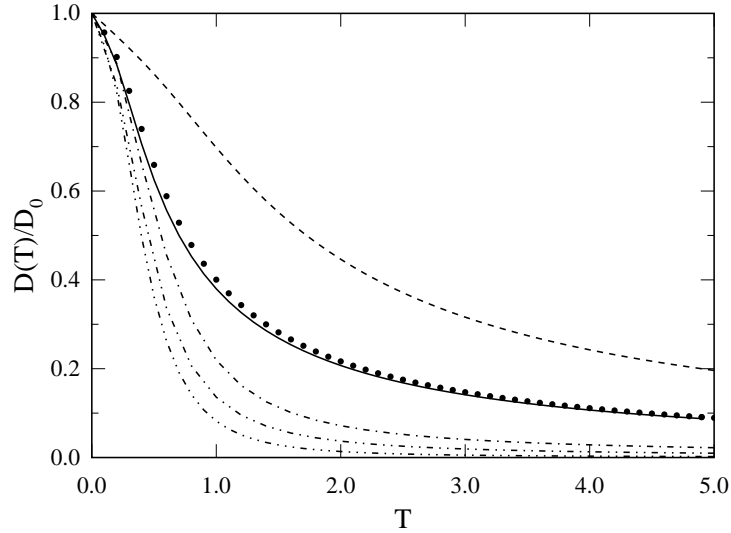}
%\hspace {0cm} {\epsfxsize 10cm \epsfysize 8cm \epsffile{figx2.ps}}
\caption{
$D(T)/D_0$ as a function of $T$ for $U/t=2$. Numerical results
are indicated by
$( -~\cdot~- )~ L=6$,
$( -~\cdot~\cdot~- )~ L=10$,
$( -~\cdot~\cdot~\cdot~- )~ L=14$
and $t_h=0.5t$; by points for $t_h=t$ and $L=18$.
The continuous line is the Bethe ansatz result and the dashed line
$D(T)/D_0$ for a free particle.}
\label{Fig. 2}
\end{figure}

Above results are clearly consistent with the conjecture \cite{czp}
that at the integrable point $t_h=t$ we are dealing with ideal
conductance (tagged particle mobility) with $D(T>0)>0$, while for $t_h
\neq t$ the transport is ``normal'' with $D(T>0)=0$ and finite
d.c. mobility $\mu_0$. A very relevant but so far open question is how
does $\mu_0$ behave close to the integrable point, where it
diverges. One would speculate on the power law divergence $\mu_0
\propto |t-t_h|^{-\gamma}$ with $\gamma>1$, which is partly supported by
numerical results \cite{cz1}.

\subsubsection{$t$-$V$ model}

In contrast to the system of a single particle in a bath the second
model, i.e. the generalized $t$-$V$ model, is a simplest model of a
homogeneous system of interacting spinless fermions
\begin{equation}
H=-t\sum_{i=1}^L (e^{i\phi}c^{\dagger}_{i+1} c_{i} + H.c.)+
V\sum_{i=1}^L n_i n_{i+1} + W \sum_{i=1}^L n_i n_{i+2}, 
\label{eq10}
\end{equation}
where the interaction is between fermions on the n.n. sites ($V$ term)
as well as next n.n. sites ($W$ term).  This Hamiltonian is integrable
using the Bethe ansatz method for $W=0$ and
non-integrable for $W\ne 0$. For $W=0$ and $V< 2t$ the ground state is
metallic, while for $V> 2t$ a charge gap opens at half filling $n=1/2$
and the system is an insulator.  For $W=0$ the $t$-$V$ model is
equivalent to the Heisenberg spin 1/2 chain as it can be shown using a
Jordan-Wigner transformation.

We studied numerically \cite{zp} various size systems with p.b.c. and
$N_e=L/2$ fermions (half-filled band).  In order to avoid an
artificial smoothing procedure of the discrete $\sigma(\omega)$
spectra of our finite size systems, it is convenient to present the
integrated and normalized intensity $I(\omega)$
\begin{equation}
I(\omega)=D^* +
\frac{2L}{\langle - T \rangle}
\int_{0^+}^{\omega} d\omega' \sigma(\omega'),\qquad
D^*=\frac{2LD}{\langle -T \rangle}.
\label{eq11}
\end{equation}

We show in Fig.~3 $I(\omega)$ and $D^*$ for a non-integrable case with
$W\neq 0$.  Indeed, as expected for a generic metallic conductor
(resistor), we find that $D^*$ scales to zero, probably exponentially
with system size $L$. At the same time $\sigma_0 > 0$ as $I(\omega)$
approaches $\omega=0$ with a finite slope.  These two results imply
that the non-integrable system indeed behaves as a normal conductor at
$T>0$.

\begin{figure}
\centering
\epsfig{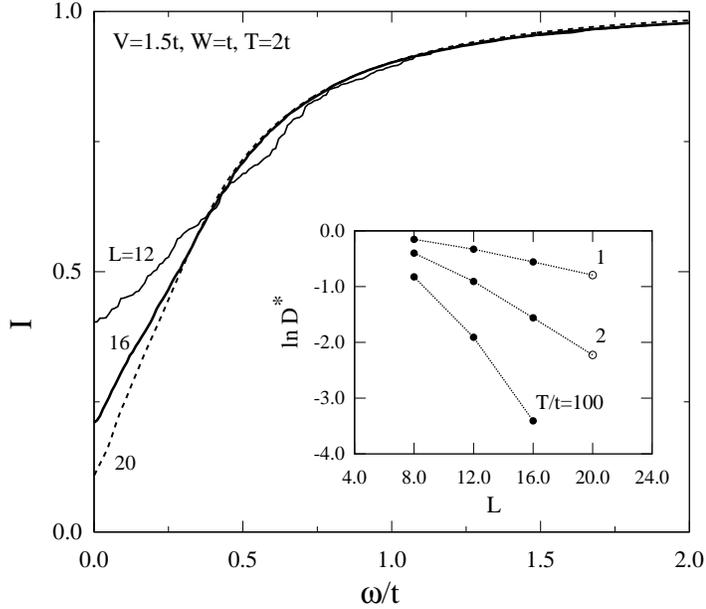}
%\hspace {1cm} {\epsfxsize 9cm \epsfysize 10cm \epsffile{figx5.ps}}
\caption{
Integrated conductivity $I(\omega)$ for $V=1.5t, W=t, T=2t$,
for $L = 12,16$ (exact diagonalization - full fines) and $L =20$
(Lanczos method - dotted lines). In the inset $ln D^*$ vs. $L$ is plotted: 
exact diagonalization (disks), $T>0$ Lanczos method (circles).}
\label{Fig. 5}
\end{figure}

In contrast to the non-integrable case the behavior of $I(\omega)$ for
an integrable system , as shown in Fig.~4, is totally different.  In
order to study the finite size dependence of the charge stiffness, we
plot in the inset $D^*$ as a function of $1/L$; the dashed lines
indicate a $3^{rd}$ order polynomial extrapolation based on the
$L=8,12,16$ site systems, suggested by the very good agreement
obtained with the $T=0$ analytical result.

\begin{figure}
\centering
\epsfig{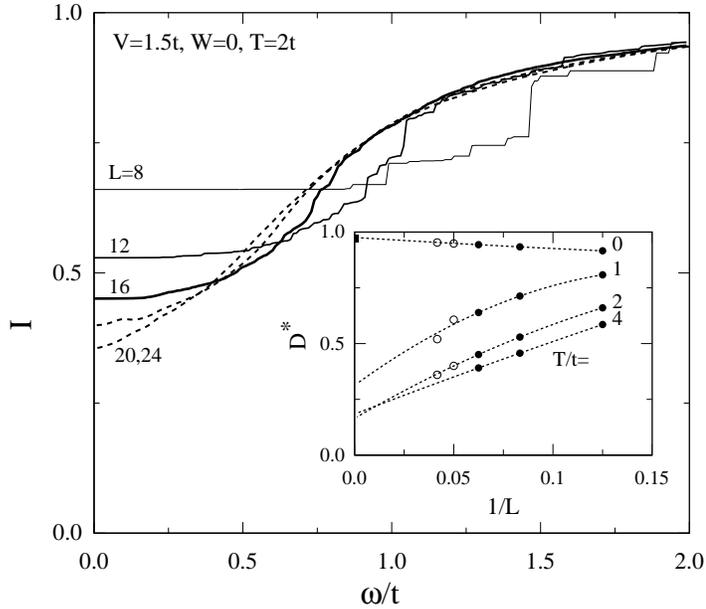}
%\hspace {1cm} {\epsfxsize 9cm \epsfysize 9.5cm \epsffile{figx6.ps}}
\caption{
Integrated conductivity $I(\omega)$ for $V=1.5t, W=0, T=2t$,
for $L = 8-16$ (exact diagonalization - full fines) and $L =20,24$
(Lanczos method - dotted lines).  Inset shows normalized charge
stiffness $D^*$ vs. $1/L$: exact diagonalization (disks), $T>0$ Lanczos
method (circles), analytical result (square) and $3^{rd}$ order
polynomial extrapolation from $L=8,12,16$ (dotted line).}
\label{Fig. 6}
\end{figure}

We find that for $L\rightarrow \infty$ the extrapolated $D^* \neq
0$. At the same time $\sigma_0=\sigma(\omega\rightarrow 0)\rightarrow
0$ as $I(\omega)$ seems to approach $\omega=0$ with zero slope.  This
behavior is reminiscent of a pseudogap.  These two results indicate
that the integrable system behaves as an {\it ideal conductor} at
$T>0$. Moreover we find that the normalized $D^*$ approaches a
nontrivial finite value $D^*_{\infty}$ in the limit $T\rightarrow
\infty$, depending on $V/t$ and on band filling, as both $D$ and
$\langle - T\rangle$ are proportional to $\beta$ in this limit.

Finally, as we mentioned in the introduction, the integrability is
expected to have an effect also in a ``Mott-Hubbard" insulating
state. In the traditional scenario, the insulator at zero temperature
is characterized by $D_0=0$ and $\sigma_0=0$; at finite temperatures
$T>0$, $D(T)$ remains zero but due to thermally activated conduction,
$\sigma_0>0$. As we will show below this scenario seems to be verified
in the generic non-integrable systems but not in the integrable ones.

The test case is the integrable $t$-$V$ model ($W=0$) at half-filling
$n=1/2$, but in the insulating regime $V>2t$.  The ground state here
is insulating characterized by $D_0=0$ and a finite charge
gap. Numerical results at finite temperatures \cite{zp}
$D^*(T>0)=I(\omega =0)$ seems also to {\it decrease} exponentially
with the system size scaling to zero for $L\rightarrow \infty$. This
precludes a possibility for ideal conductivity at $T>0$. Furthermore
$I(\omega)$ seems to approach $\omega=0$ with a zero slope, showing a
depletion of weight within a low frequency region of order $\omega <
t$.  These are characteristics of an {\it ideal insulator}, not
conducting even at high temperatures $T \gg \Delta_0$.

In contrast, non-integrable systems of roughly the same charge gap
exhibit a qualitatively different behavior. $I(\omega)$ approaches
$\omega=0$ with finite (although small) slope, consistent with a small
static conductivity $\sigma_0>0$, consistent with the conduction due
to electron - hole pairs, thermally activated over the charge gap.

\subsection{Transport and conservation laws}

The integrable quantum many-body systems, as the Heisenberg spin 1/2
chain or the one dimensional Hubbard model, are characterized by a
macroscopic number of conserved quantities. It is natural to think
that the singular transport behavior is related to the existence of
such conserved quantities.  A set of them is represented by local
operators $Q_n$, commuting with each other $[Q_n,Q_m]=0$ and with the
Hamiltonian, $[Q_n,H]=0$.  The index $n$ indicates that the operator
$Q_n$ is of the form $Q_n=\sum_{l=1}^L q_l^n$, where $q_l^n$ are local
operators involving $n$ sites around site $l$, on a lattice of $L$
sites.

The physical content of $Q_n$ is not clear in general.  Nevertheless,
for different models discussed below, the first nontrivial quantity
$Q_3$ ($Q_2$ is the Hamiltonian) has a simple physical
significance. It is equal or closely related to the energy current
operator $j^E$. Through its coupling to the current operator $j$ it
gives an alternative explanation for the singular $D(T>0)>0$.

The time decay of correlations is related to conserved quantities in a 
Hamiltonian system by an inequality introduced by Mazur\cite{m,su}
\begin{equation}
\lim_{T\rightarrow \infty} \frac{1}{T} \int_0^T <A(t)A> dt \geq \sum_n 
\frac{<A Q_n>^2}{<Q_n^2>} ,
\label{mazur}
\end{equation}
which is based on a general positivity of spectra which hold for
stationary processes. Here, operators are assumed hermitian,
$A^{\dagger}=A$, and $< >$ denotes the thermodynamic average. The left
hand side can be writen as a sum of a time-independent constant
$\tilde D$ and a time-dependent (decaying) function $C(t)$, expressed
in terms of eigenstates,
\begin{eqnarray}
<A(t)A>&=&\sum_{n,m (\epsilon_m=\epsilon_n)} p_n |<n|A|m>|^2 + 
\sum_{n,m (\epsilon_m\neq \epsilon_n)} p_n 
|<n|A|m>|^2 e^{i (\epsilon_n - \epsilon_m) t} \nonumber \\
&=&\tilde D + C(t). \label{maz2}
\end{eqnarray}
For time correlations $<A(t)A>$ with a non-singular low frequency
behavior the second term $\lim_{T\rightarrow \infty} \frac{1}{T}
\int_0^T C(t) dt$ goes to zero and therefore
\begin{equation}
\tilde D \geq \sum_n \frac{<A Q_n>^2}{<Q_n^2>}. \label{maz3}
\end{equation}

We can use this inequality for the analysis of $\sigma(\omega)$
related to the current-current correlation $<j(t)j>$.  Charge
stiffness $D$, Eq.(\ref{eq7}), in this case is directly related to
$\tilde D$, and we get
\begin{equation}
D \geq \frac{\beta}{2L} \sum_n \frac{<J Q_n>^2}{<Q_n^2>} .
\label{eqd}
\end{equation}
In this derivation we assume that the regular part
$\sigma_{reg}(\omega)$ shows a non-singular behavior at low
frequencies so that the contribution from $C(t)$ in (\ref{mazur})
vanishes. The latter is consistent with numerical simulations showing
for integrable systems a pseudogap at small $\omega$ and so a
vanishing regular part $\sigma_{reg}(\omega\rightarrow 0)$.

In general it is difficult to evaluate the right hand side of the
inequality (\ref{mazur}) involving the ``overlap" $<A Q_n>$. However,
in some examples presented below, this correlation can easily be
evaluated in the case of a grand canonical trace over states, in the
thermodynamic limit and for $\beta \rightarrow 0$.

Before studying concrete models we construct the energy current
operator $j^E$ for tight-binding models with n.n. hopping only,
where the Hamiltonian can be presented as a sum
\begin{equation}
H=\sum_{i=1}^L h_{i,i+1}. \label{H}
\end{equation}
where $h_{i,i+1}$ are local energy operators acting on a bond.
$H$ is a conserved quantity, hence the time evolution of the 
local $h_{i,i+1}$  can be writen as the divergence of
energy current operator $j^E_i$:
\begin{equation}
\frac{\partial h_{i,i+1}}{\partial t}= i [H,h_{i,i+1}]= 
- ( j^E_{i+1}- j^E_i) ,
\label{je}
\end{equation}
and we get for the (total) energy current
\begin{equation}
j^E=\sum_{i=1}^L j^E_i= -i \sum_{i=1}^L [h_{i-1,i},h_{i,i+1}].
\end{equation}

\subsubsection{$t$-$V$ model: Heisenberg model}

$t$-$V$ model is through the through a Jordan-Wigner transformation
equaivalent to the anisotropic Heisenberg spin ($S=1/2$) Hamiltonian,
provided that $J_x=J_y$,
\begin{equation}
H=\sum_{i=1}^L \bigl [J_x S_i^x S_{i+1}^x + 
J_y S_i^y S_{i+1}^y + J_z S_i^z S_{i+1}^z \bigr]. \label{heis}
\end{equation}
Here the local energy current operators $j^E_i$ can be expressed as
\begin{equation}
j^E_i=J_x J_y (xzy - yzx)_{i-1,i+1} + J_y J_z (yxz - zxy)_{i-1,i+1} 
+ J_z J_x (zyx - xyz)_{i-1,i+1} ,
\label{jeheis} \end{equation}
with $(\alpha\beta\gamma - \gamma\beta\alpha)_{i-1,i+1}=
S_{i-1}^{\alpha} S_i^{\beta} S_{i+1}^{\gamma} - 
S_{i-1}^{\gamma} S_i^{\beta} S_{i+1}^{\alpha}$. 
 
It is straigthforward to verify that the global energy current
operator $j^E$ commutes with the $H$ (\ref{heis}) and coincides with
$Q_3$. Consistent with the notation $Q_2$ for the Hamiltonian is the
sum of local operators involving two sites $q_2=h_{i,i+1}$,
$q^3_i=j^E_i$ involves three neighboring sites etc.

The vanishing commutator $[j^E,H]=0$ implies also that the energy
current time correlations are independent of time.  The non-decaying
of the energy current then leads evidently to a {\it singular thermal
conductivity} $\kappa$ related to the $<j^E(t)j^E>$ correlation as
well as a {\it singular thermopower} related to $<j^E(t) j>$.

In the language of the fermionic $t$-$V$  model $j^E$ is given by
\begin{equation}
j^E_i = t^2(i c_{i+1}^{\dagger} c_{i-1} + h.c.)+ V
j_{i,i+1}(n_{i-1}+n_{i+2}-1), \label{jqtv}
\end{equation}
where $j_{i,i+1}=t(i c_{i+1}^{\dagger} c_{i} + h.c.)$ is the local
particle current.  From (\ref{eqd}) one can get an explicit expression
for $\beta\rightarrow 0$, where the leading order of the high-$T$
expansion of $\langle j^Ej\rangle$ and $\langle jj\rangle$ yields
\begin{equation}
D\geq \frac{\beta}{2} \frac{2 V^2 n (1-n) (2n -1)^2} 
{1+V^2(2n^2-2n+1)}. \label{dtv}
\end{equation}
We note that outside the half-filling $n \neq 1/2$, $D>0$ implies
again an ideal conductivity as shown before using the argument of a
relation to the level dynamics. For $n=1/2$ the inequality is however
insufficient for proving that $D$ is nonzero.  Due to the
electron-hole symmetry, this remains true even if we consider all
higher order conserved quantities $Q_n$. This evidently presents an
open problem, since the inequality seems to be exhausted neglecting
possible nonlocal conserved quantities.

\subsubsection {Hubbard model}

As above one can define a local energy operator for the Hubbard model,
\begin{eqnarray}
h_{i,i+1}=&&-t\sum_{s} (c_{is}^{\dagger} c_{i+1, s} + h.c.) +
\nonumber \\ &&+ \frac{U}{2} \bigl[
(n_{i\uparrow}-\frac{1}{2})(n_{i\downarrow}-\frac{1}{2})+
(n_{i+1\uparrow}-\frac{1}{2})(n_{i+1\downarrow}-\frac{1}{2}) \bigr],
\label{eq101}
\end{eqnarray}
and 
\begin{equation}
j^E_i=\sum_{s} t^2(i c_{i+1,s}^{\dagger} c_{i-1,s} + h.c.)-
\frac{U}{2} (j_{i-1,i,s}+j_{i,i+1,s}) (n_{i,-s}-\frac{1}{2}) ,
\label{eq111}
\end{equation}
where $j_{i,i+1,s}=(-t)(-i c_{i+1,s}^{\dagger} c_{is} + h.c.)$ is the
particle current. By comparing the expression for $j^E$ to $Q_3$, we
find that they coincide when the factor $U/2$ in (\ref{eq111}) is
replaced by $U$. One can verify that the energy current $j^E$ has a
finite overlap $<j^E Q_3>$, with the conserved quantity $Q_3$.  We
therefore conclude that although the energy current correlations are
time dependent still they do not decay to zero so that the thermal
transport coefficients are singular.

Using again (\ref{eqd}) we find for the charge stiffness $D$ of the 
Hubbard model, again at $\beta\rightarrow 0$,
\begin{equation}
D \geq \frac{\beta}{2} U^2 \frac {\sum_{s}n_{s}(1-n_{s})(2n_{-s}-1)}
{2\sum_{s} n_{s}(1-n_{s})[1+(U/2)^2(2n_{-s}^2- 2n_{-s}+1)]} ,
\label{dhubb}
\end{equation}
where $n_{s}$ are densities of $s=\uparrow,\downarrow$ fermions.
Again, we get $D(T)>0$ outside half filling $n_{s}\neq 1/2$.  On the
other hand $D(T)=0$ for $n_{s}=1/2$. In the latter case the validity
or deficiency of the result is not so clear, since the insulating
$D=0$ at half filling is not unexpected and there are so far also no
reliable analytical or numerical result which could be confronted with
the one following from the Mazur inequality.

\section{Numerical methods}

Let us turn to the discussion of higher-dimensional (as well as
nonintegrable 1D) correlated systems which behave as normal metals
with finite d.c. transport quantities, e.g. the resistivity $\rho(T)$.
It is evident that the evaluation of $\sigma_0(T)$ via the linear
response theory requires the method at $T>0$.

The absense of well-controlled analytical approaches to models of
strongly correlated electrons has stimulated the development of
computational methods. Conceptually the simplest is the exact
diagonalization method of small systems \cite{dago}.  In models of
correlated electrons, however, one is dealing with the number of
quantum basis states $N_{st}$ which grows exponentially with the size
of the system. Within the Hubbard model there are 4 basis states for
each lattice site, therefore the number of basis states in the
$N$-site system is $N_{st} \propto 4^N$. In the $t$-$J$ model $N_{st}$
still grows as $\propto 3^N$, and within the $t$-$V$ (or Heisenberg)
model as $\propto 2^N$.  In the exact diagonalization of such systems
one is therefore representing operators with matrices $N_{st}\times
N_{st}$, which become large already for very modest values of $N$.

One straightforward approach to dynamics at $T>0$ is to perform full
diagonalization of small system, calculating all eigenstates and
eigenfunctions $\epsilon_n, |n\rangle$, required to evaluate matrix
elements, e.g.  $\langle n| j |m \rangle$. At the present status of
computers we are restricted to diagonalization of matrices with
$N_{st} <3000$, so that reachable systems are $N\leq 16$, $N\leq 12$,
$N \leq 8$ for the $t$-$V$ model, the $t$-$J$ model and the Hubbard
model, respectively.

\subsection{Lanczos method}

The helpful circumstance is that for most interesting operators and
lattice models only a small proportion of matrix elements is nonzero
within the local basis.  Then, the operators can be represented by
sparse matrices with $N_{st}$ rows and at most $f(N)$ nonzero elements
in each row. In this way memory requirements are relaxed and matrices
up to $N_{st} \sim 10^8$ are considered in recent applications.
Finding eigenvalues and eigenvectors of such large matrices is not
possible with standard algorithms performing the full
diagonalization. One must instead resort to power algorithms, among
which the Lanczos algorithm is one of the most widely known
\cite{part}.  It is mainly used for the calculation of the ground
state of a finite system, i.e. the ground-state energy $E_0$ and
wavefunction $|\Psi_0\rangle $.

The Lanczos algorithm \cite{part} starts with a normalized random
vector $|\phi_0\rangle$ in the vector space in which the Hamiltonian
operator $H$ is defined. $H$ is applied to $|\phi_0\rangle$ and the
resulting vector is split in components parallel to $|\phi_0\rangle$,
and $|\phi_1\rangle$ orthogonal to it, respectively,
\begin{equation}
H|\phi_0\rangle=a_0 |\phi_0\rangle + b_1|\phi_1\rangle.
\label{fl1}
\end{equation}
Since $H$ is Hermitian, $a_0=\langle\phi_0|H|\phi_0\rangle$ is real,
while the phase of $|\phi_1\rangle$ can be chosen so that $b_1$ is
also real.  In the next step $H$ is applied to $|\phi_1\rangle$,
\begin{equation}
H|\phi_1\rangle=b_1'|\phi_0\rangle +a_1 |\phi_1\rangle + b_2|\phi_2\rangle,
\label{fl2}
\end{equation}
where $|\phi_2\rangle$ is orthogonal to $|\phi_0\rangle$ and
$|\phi_1\rangle$. It follows also $b_1'=\langle\phi_0|H|\phi_1\rangle
= b_1$. Proceeding with the iteration one gets in $i$ steps
\begin{equation}
H|\phi_i\rangle=b_i|\phi_{i-1}\rangle +a_i |\phi_i\rangle + 
b_{i+1}|\phi_{i+1}\rangle,\qquad 1\leq i \leq M. \label{fl3}
\end{equation}

By stopping the iteration at $i=M$ and putting the last coefficient
$b_{M+1}=0$, the Hamiltonian can be represented in the basis of
orthogonal Lanczos functions $|\phi_i\rangle$ as the tridiagonal
matrix $H_M$ with diagonal elements $a_i$ with $i=0\ldots M$, and
offdiagonal ones $b_i$ with $i=1\ldots M$.  Such a matrix is easily
diagonalized using standard numerical routines to obtain approximate
eigenvalues $\epsilon_j$ and the corresponding orthonormal
eigenvectors $|\psi_j\rangle$,
\begin{equation}
|\psi_j\rangle=\sum_{i=0}^M v_{ji} |\phi_i\rangle,\;\;\;j=0\ldots M.
\label{fl5}
\end{equation}

It is important to realize that $|\psi_j\rangle$ are (in general) not
exact eigenfunctions of $H$, but show a remainder
\begin{equation}
H|\psi_j\rangle-\epsilon_j|\psi_j\rangle= b_{M+1}v_{jM}|\phi_{M+1}\rangle.
\label{fl6}
\end{equation}
On the other hand it is evident from the diagonalization of $H_M$ that
matrix elements
\begin{equation}
\langle\psi_i|H|\psi_j\rangle=\epsilon_j\delta_{ij},\;\;\;i,j=0\ldots M
\label{fl7}
\end{equation}
are exact, without restriction to the subspace $L_M$.

The identity (\ref{fl7}) already shows the usefulness of the Lanczos
method for the calculation of particular matrix elements. As an aid in a
further discussion of the Lanczos method we consider the
calculation of a matrix element
\begin{equation}
W_{kl}=\langle n|H^k B H^l A|n\rangle, \label{fe1}
\end{equation}
where $|n\rangle$ is an arbitrary normalized vector, and $A, B$ are
general operators. One can calculate this expression exactly by
performing two Lanczos procedures with $M=\max(k,l)$ steps. The first
one, starting with the vector $|\phi_0\rangle=|n\rangle$, produces the
subspace $L_M=\{|\phi_j\rangle,\;j=0\ldots M\}$ along with approximate
eigenvectors $|\psi_j\rangle$ and eigenvalues $\epsilon_j$. The second
Lanczos procedure is started with the normalized vector
\begin{equation}
|\tilde\phi_0\rangle=A|\phi_0\rangle/\sqrt{\langle\phi_0| A^\dagger
A|\phi_0\rangle}, \label{fe2}
\end{equation}
and results in the subspace $\tilde
L_M=\{|\tilde\phi_j\rangle,\;j=0\ldots M\}$ with approximate
$ |\tilde\psi_j\rangle$ and $\tilde \epsilon_j$. We can now define projectors
\begin{equation}
P_m=\sum_{i=0}^m|\phi_i\rangle\langle\phi_i|,\;\;
\tilde P_m=\sum_{i=0}^m|\tilde\phi_i\rangle\langle\tilde\phi_i|, \label{fe3}
\end{equation}
which for $m=M$ can also be expressed as
\begin{equation}
P_M=\sum_{i=0}^M|\psi_i\rangle\langle\psi_i|,\;\;
\tilde P_M=\sum_{i=0}^M|\tilde\psi_i\rangle\langle\tilde\psi_i|.
\label{fe4}
\end{equation}
By taking into account definitions (\ref{fe3}), (\ref{fe4}) we show
that
\begin{equation}
H P_m=P_{m+1}HP_m=P_M H P_m, \qquad m<M. \label{fe5}
\end{equation}
Since in addition $|n\rangle=|\phi_0\rangle=P_0|\phi_0\rangle$
and $A|n\rangle\propto|\tilde\phi_0\rangle=P_0|\tilde\phi_0\rangle$,
by successive use of the first equality in (\ref{fe5}) we 
arrive at 
\begin{equation}
W_{kl}=
\langle \phi_0|P_MHP_MH\ldots HP_MB\tilde P_MH\ldots\tilde P_M
H\tilde P_M A|\phi_0\rangle. \label{fe7}
\end{equation}
We note that the necessary condition for the equation (\ref{fe7}) is
$M\ge k,l$. We finally expand the projectors according to expressions
(\ref{fe4}) and take into account the orthonormality relation
(\ref{fl7}) for matrix elements, and get
\begin{eqnarray}
W_{kl}&=&
\sum_{i_0=0}^M\ldots\sum_{i_k=0}^M\sum_{j_0=0}^M\ldots\sum_{j_l=0}^M
\langle\phi_0|\psi_{i_0}\rangle\langle\psi_{i_0}|
H|\psi_{i_1}\rangle\ldots \langle\psi_{i_{k-1}}|H|\psi_{i_k}\rangle 
\nonumber \\
& &\times \langle\psi_{i_k}|B|\tilde\psi_{j_l}\rangle
\langle\tilde\psi_{j_l}|
H|\tilde\psi_{j_{l-1}}\rangle\ldots \langle \tilde\psi_{j_1}|H
|\tilde\psi_{j_0}\rangle
\langle\tilde\psi_{j_0}|A|\phi_0\rangle =\nonumber \\
&=&
\sum_{i=0}^M\sum_{j=0}^M\langle\phi_0|\psi_{i}\rangle\langle\psi_{i}|
B|\tilde\psi_{j}\rangle\langle\tilde\psi_{j}|A|\phi_0\rangle
(\epsilon_i)^k (\tilde \epsilon_j)^l. \label{fe8}
\end{eqnarray}
We have thus expressed the desired quantity in terms of the
Lanczos (approximate) eigenvectors and eigenvalues alone.

Within the Lanczos algorithm the extreme (smallest and largest)
eigenvalues $\epsilon_i$, along with their corresponding
$|\psi_i\rangle$, are rapidly converging to exact eigenvalues $E_i$
and eigenvectors $|\Psi_i\rangle$. It is quite characteristic that
usually (for nondegenerate states) $M=30-60 \ll N_{st}$ is sufficient
to achieve the convergence to the machine precision of the ground
state energy $E_0$ and the wavefunction $|\Psi_0\rangle$, from which
various static and dynamical correlation functions at $T=0$ can be
evaluated.

After $|\Psi_0\rangle$ is obtained, the g.s. dynamic correlation
functions can be calculated within the same framework
\cite{hayd}. Let us consider the autocorrelation function
\begin{equation}
C(t)=-i\langle\Psi_0|A^\dagger(t)A|\Psi_0\rangle
=-i\langle\Psi_0|A^\dagger e^{i(E_0-H)t}A|\Psi_0\rangle
\label{fd1}
\end{equation}
with the transform, 
\begin{equation}
\tilde C(\omega)=\int_0^\infty dt e^{i\omega^+t}C(t) =\langle\Psi_0|
A^\dagger \frac{1}{\omega^+ +E_0-H}A|\Psi_0\rangle. \label{fd2}
\end{equation}
To calculate $\tilde C(\omega)$, one has to run the second Lanczos
procedure starting with the normalized function
$|\tilde\phi_0\rangle$, equation (\ref{fe2}).  The matrix for $H$ in
the new basis $\tilde L_M$, with elements
$\langle\tilde\phi_i|H|\tilde\phi_j\rangle=[\tilde H_M]_{ij}$, is
again a tridiagonal one with $\tilde a_i$ and $\tilde b_i$ elements,
respectively.  Terminating the Lanczos procedure at given $M$, one can
evaluate the $\tilde C(\omega)$ as a resolvent of the $\tilde H_M$
matrix which can be expressed in the continued-fraction form
\cite{hayd},
\begin{equation}
\tilde C(\omega)=\frac{\langle\Psi_0|A^\dagger A|\Psi_0\rangle}
{\omega^+ +E_0-\tilde a_0-{\displaystyle\frac{\tilde b_1^2}
{\omega^+ +E_0-\tilde a_1-{\displaystyle\frac{\tilde b_2^2}
{\omega^+ +E_0-\tilde a_2-\ldots}}}}}\;, \label{fd3}
\end{equation}
terminating with $\tilde b_{M+1}=0$, although other termination functions
have also been employed.

The spectral function $C(\omega)=-(1/\pi) {\rm Im} \tilde C(\omega)$
is characterized by frequency moments,
\begin{equation}
\mu_l=\int_{-\infty}^\infty \omega^l C(\omega) d\omega
=\langle\Psi_0|A^\dagger(H-E_0)^lA|\Psi_0\rangle,
\label{fd4}
\end{equation}
which are particular cases of the expression (\ref{fe1}) for
$B=A^\dagger$, $k=0$, and $|n\rangle=|\Psi_0\rangle$.  Using the
equation (\ref{fe8}) we can express $\mu_l$ for $l\le M$ in terms of
Lanczos quantities alone
\begin{equation}
\mu_l=\sum_{j=0}^M 
\langle\Psi_0|A^\dagger|\tilde\psi_j\rangle
\langle\tilde\psi_j|A|\Psi_0\rangle
(\tilde \epsilon_j-E_0)^l. \label{fd5}
\end{equation}
Hence moments $\mu_{l<M}$ are exact for given $|\Psi_0\rangle$.

\subsection{Finite temperature Lanczos method}

The novel method for $T>0$ \cite{jp,jprev} is based on the application
of the Lanczos iteration, reproducing correctly high-$T$ and
large-$\omega$ series. The method is then combined with the reduction
of the full thermodynamic trace to the random sampling.

We first consider the expectation value of the operator $A$ in the
canonical ensemble
\begin{equation}
\langle A\rangle=\sum_{n=1}^{N_{st}}\langle n|e^{-\beta H}A|n\rangle
\biggm/
\sum_{n=1}^{N_{st}}\langle n|e^{-\beta H}|n\rangle.
\label{fh1}
\end{equation}
A straightforward calculation of $\langle A\rangle$ requires the
knowledge of all eigenstates $|\Psi_n\rangle$ and corresponding
energies $E_n$, obtained by the full diagonalization of $H$,
computationally accessible only for $N_{st} < 3000$. Instead let us
perform the high-temperature expansion of the exponential ${\rm
exp}(-\beta H)$,
\begin{eqnarray}
\langle A\rangle &=& Z^{-1}
\sum_{n=1}^{N_{st}}\sum_{k=0}^\infty
\frac{(-\beta)^k}{k!}\langle n|H^kA|n\rangle, \nonumber\\
Z&=&\sum_{n=1}^{N_{st}}\sum_{k=0}^\infty
\frac{(-\beta)^k}{k!}\langle n|H^k|n\rangle. \label{fh2}
\end{eqnarray}
Terms in the expansion $\langle n|H^k A|n\rangle$ can be calculated
exactly using the Lanczos procedure with $M \geq k$ steps and with
$|\phi^n_0\rangle=|n\rangle$ as a starting function, since this is a
special case of the expression (\ref{fe1}). Using the relation
(\ref{fe8}) with $l=0$ and $B=1$, we get
\begin{equation}
\langle n|H^k A|n\rangle=
\sum_{i=0}^M\langle n|\psi^n_{i}\rangle\langle\psi^n_{i}|
A|n\rangle (\epsilon^n_i)^k. \label{fh3}
\end{equation}
Working in a restricted basis $k\leq M$, we can insert the expression
(\ref{fh3}) into sums (\ref{fh2}), extending them to $k >M$.  The
final result can be expressed as
\begin{eqnarray}
\langle A \rangle &\approx& Z^{-1}\sum_{n=1}^{N_{st}}\sum_{i=0}^M
e^{-\beta \epsilon^n_i}\langle n|\psi^n_i\rangle\langle\psi^n_i|A|n
\rangle, \nonumber \\
Z &\approx& \sum_{n=1}^{N_{st}}\sum_{i=0}^M e^{-\beta
\epsilon^n_i}\langle n|\psi^n_i\rangle\langle\psi^n_i|n
\rangle, \label{fh4}
\end{eqnarray}
and the error of the approximation is of the order of 
$\beta^{M+1}$.

Evidently, within a finite system the expression (\ref{fh4}), expanded
as a series in $\beta$, reproduces exactly the high-$T$ expansion
series to the order $M$. In addition, in contrast to the usual series
expansion, it becomes (remains) exact also for $T\to 0$. Let us assume
for simplicity that the ground state $|\Psi_0\rangle$ is
nondegenerate.  For initial states $|n\rangle$ not orthogonal to
$|\Psi_0\rangle$, already at modest $M <60 $ the lowest function
$|\psi^n_0\rangle$ converges to $|\Psi_0\rangle$. Eq.(\ref{fh4}) for
$\beta \to \infty$ then gives,
\begin{eqnarray}
\langle A\rangle
&=&\sum_{n=1}^{N_{st}}
\langle n|\Psi_0\rangle\langle\Psi_0|A|n\rangle\bigg/
\sum_{n=1}^{N_{st}}\langle n|\Psi_0\rangle\langle\Psi_0|n\rangle =\nonumber \\
&=&\langle\Psi_0|A|\Psi_0\rangle/\langle\Psi_0|\Psi_0\rangle,
\label{fh5}
\end{eqnarray}
where we have taken into account the completeness of the set
$|n\rangle$.  Obtained result is just the usual ground state
expectation value of an operator.

In order to calculate dynamical quantities, the high-$T$ expansion
must be supplemented by the high-frequency (short-time) expansion.
The goal is to calculate the dynamical correlation function at $T>0$,
\begin{equation}
\langle B(t)A\rangle={\rm Tr}\left[e^{-\beta H}e^{iHt}Be^{-iHt}A\right]/
{\rm Tr}~e^{-\beta H}. \label{ff1}
\end{equation}
Expressing the trace explicitly and expanding the exponentials, we get
\begin{equation}
\langle B(t)A\rangle = Z^{-1}
\sum_{n=1}^{N_{st}}\sum_{k=0}^\infty\sum_{l=0}^\infty
\frac{(-\beta+it)^k}{k!}\frac{(-it)^l}{l!}
\langle n|H^kBH^lA|n\rangle. \label{ff2}
\end{equation}
Expansion coefficients in equation (\ref{ff2}) can be again obtained
via the Lanczos method.  Performing two
Lanczos iterations with $M$ steps, started with normalized
$|\phi^n_0\rangle=|n\rangle$ and $|\tilde\phi^n_0\rangle \propto
A|n\rangle$, respectively, we calculate coefficients $W_{kl}$
following the equation (\ref{fe8}), while $Z$ is approximated by the
static expression (\ref{fh4}). Extending and resumming series in $k$
and $l$ into exponentials, we get
\begin{equation}
\langle B(t)A\rangle \approx Z^{-1}
\sum_{n=1}^{N_{st}}
\sum_{i=0}^M\sum_{j=0}^M e^{-\beta \epsilon^n_i}
e^{it(\epsilon^n_i-\tilde \epsilon^n_j)}
\langle n|\psi^n_{i}\rangle\langle
\psi^n_{i}|B|\tilde\psi^n_{j}\rangle\langle\tilde\psi^n_{j}|A
|n\rangle.
\label{ff3}
\end{equation}

The computation of static quantities (\ref{fh4}) and dynamical ones
(\ref{ff3}) still involves the summation over the complete set of
$N_{st}$ states $|n\rangle$, which is not feasible in practice.  To
obtain a useful method, one further approximation must be made which
replaces the full summation by a partial one over a much smaller set
of random states. Such an approximation, analogous to Monte Carlo
methods, is of course hard to justify rigorously, nevertheless we can
estimate the errors involved.

We consider the expectation value $\langle A \rangle$ at $T>0$, as
defined by the expression (\ref{fh1}).  Instead of the whole sum in
equation (\ref{fh1}) we first evaluate only one element with respect
to a random state $|r\rangle$, which is a linear combination of basis
states
\begin{equation}
|r\rangle=\sum_{n=1}^{N_{st}}\beta_{rn}|n\rangle, \label{fr1}
\end{equation}
i.e. $\beta_{rn}$ are assumed to be distributed  randomly.
Let us discuss then the random quantity
\begin{equation}
\tilde A_r=\langle r|e^{-\beta H}A|r\rangle/
\langle r|e^{-\beta H}|r\rangle 
\end{equation}
We express $|\beta_{rn}|^2=1/N_{st}+\delta_{rn}$ and assume that
random deviations $\delta_{rn}$ are not correlated with matrix
elements $\langle n|e^{-\beta H}|n\rangle=Z_n$ and $\langle
n|e^{-\beta H}A|n\rangle=Z_n A_n$.  Then it is easy to show that
$\tilde A_r$ is close to $\langle A\rangle$, and the statistical
deviation is related to the effective number of terms $\tilde Z$ in
the thermodynamic sum, i.e.
\begin{equation}
\tilde A_r = \langle A\rangle +{\cal O}(1/\sqrt{\bar Z}),\qquad
\bar Z=e^{\beta E_0}\sum_n Z_n \sim {\rm Tr}e^{-\beta (H-E_0)}.
\label{fr4}
\end{equation}
Note that for $T\to \infty$ we have $\bar Z\to N_{st}$ and therefore
at large $N_{st}$ a close estimate of the average (\ref{fr4}) can be
obtained from a single random state. On the other hand, at finite
$T<\infty$ the statistical error of $\tilde A_r$ increases with
decreasing $\bar Z$. In the FTLM we replace the full summation in the
expression (\ref{fh1}) with a restricted one over several random
vectors $|r\rangle$, $r=1,R$ and the statistical error is even
reduced,
\begin{equation}
\tilde A = \langle A\rangle +{\cal O}(1/\sqrt{R\bar Z}).
\label{fr6}
\end{equation}

We introduce and justify the finite-temperature Lanczos method as a
method to calculate $T>0$ properties on small systems, but as well as
a controlled approach to results at $T \to 0$ in the thermodynamic
limit. We also argue that choosing appropriate $M$, and the sampling
$R$ one can reproduce exact results to prescribed precision on a given
system.  The well known deficiency of the exact diagonalization
methods is however the smallness of available lattices, hence it is
important to understand the finite size effects and their role at
$T>0$. We claim that generally $T>0$ reduces finite size effects.
This is related to the fact that at $T=0$ both static and dynamical
quantities are calculated only from one wavefunction $|\Psi_0\rangle$,
which can be quite dependent on the size and on the shape of the
system.  $T>0$ introduces the thermodynamic averaging over a larger
number of eigenstates. This reduces directly finite-size effects for
static quantities, whereas for dynamical quantities spectra become
denser.

\begin{figure}
\centering
\epsfig{file=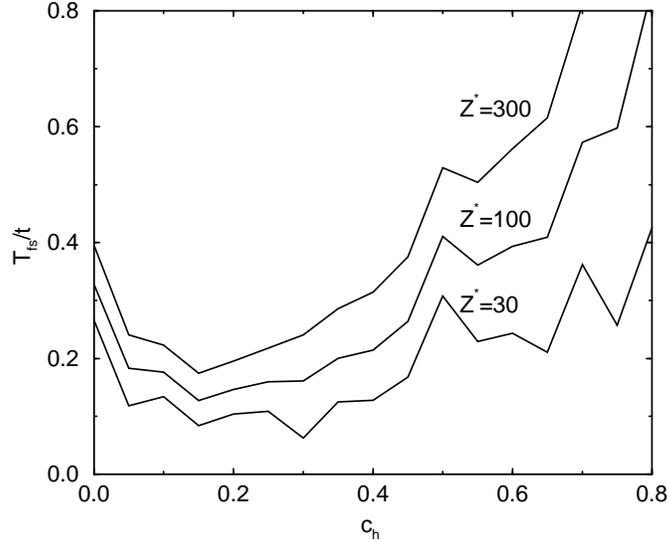,height=10cm,angle=-90}
\caption{
The variation of $T_{fs}$ with doping $c_h$ in the $t$-$J$ on $N=18$
sites with $J/t=~0.3$. Curves correspond to different choices of
$Z^*=\bar Z(T_{fs})$.  } \label{3.6}
\end{figure}

The effect of $T>0$ can be expressed also in another way.  There are
several characteristic length scales in the system of correlated
electrons, e.g. the antiferromagnetic correlation length $\xi$, the
transport mean free path $l_s$, etc.  These lengths decrease with
increasing $T$ and results for related quantities have a macroscopic
relevance provided that the lengths become shorter than the system
size, e.g. $l_s < L$ where $L$ is the linear size of the system. This
happens for particular $T> T_s$, where clearly $T_s$ depends also on
the quantity considered. 

As a criterion for finite size effects we use the characteristic
finite-size temperature $T_{fs}$. It is chosen so that in a given
system the thermodynamic sum in Eq.(\ref{fr4}) is appreciable,
i.e. $\bar Z(T_{fs})=Z^* \gg 1$. The finite-temperature Lanczos method
is thus best suited just for quantum many-body systems with a large
degeneracy of states, i.e. large $\bar Z$ at low $T$.  This is the
case with doped antiferromagnet and the $t$-$J$ model in the strong correlation
regime $J<t$. To be concrete we present in Fig.~\ref{3.6} the
variation of $T_{fs}$ with the doping $c_h=N_h/N$, as calculated from
the system of $N=18$ sites and $J/t=0.3$.  It is indicative that
$T_{fs}$ reaches the minimum for intermediate (optimum) doping $c_h
=c^*_h \sim 0.15$. Away from such optimum case $T_{fs}$ is larger.  We
claim that small $T_{fs}$ and related large degeneracy of low-lying
states are the essential features of strongly correlated system in
their most challenging regime (so called underdoped and optimally
doped regime of cuprates), being a sign of a novel quantum
frustration.  This is consistent with the experimental results on
cuprates, where the optimal doping with respect to $T_c$ coincides
with largest degeneracy, i.e. electronic entropy density $s$.
         
\section{$t$-$J$ model vs. cuprates}

The main motivation to understand transport in strongly correlated
systems comes from the well established anomalous properties of
high-$T_c$ cuprates, whereby in our context we consider only the
''normal'' metallic state. The prominent example is nearly linear
in-plane resistivity $\rho \propto T$ in the normal state
\cite{imad}. It is however an experimental fact that such a behaviour
is restricted to the optimum doping regime, while deviations from
linearity appear both in the underdoped and overdoped regimes, being
still universal for a number of materials with a similar doping.  The
d.c. resistivity $\rho$ is intimately related to the optical
conductivity $\sigma(\omega)$, which has been also extensively studied
and shows in the normal state the unusual non-Drude behaviour fitted
with the marginal Fermi-liquid \cite{varm} behaviour $\tau \sim
2\pi\lambda (\omega + \xi T)$.

There is a lot of evidence that the anomalous electrical transport in
cuprates is mainly due to strong correlations. Moreover, it seems that
at least within the normal metallic state the physics of cuprates can
be well captured within the single-band $t$-$J$ model, the simplest
model which contains the interplay of the charge propagation and spin
(antiferromagnetic) exchange, 
\begin{equation}
H=-t\sum_{\langle ij\rangle  s}(\tilde{c}^\dagger_{js}\tilde{c}_{is}+
{\rm H.c.})+J\sum_{\langle ij\rangle} ({\bf S}_i\cdot {\bf S}_j -
\frac{1}{4} n_i n_j). \label{cm1}
\end{equation}
Here ${\bf S}_i$ are the local spin operators interacting with the
exchange parameter $J$.  Due to the strong on-site repulsion the
double occupion of sites is explicitly forbidden and we are dealing
with projected fermion operators $\tilde{c}_{is}=c_{is}(1-n_{i,-s})$.

Due to small number of local degrees of freedom ($N_i=3$) the $t$-$J$
model is particularly convenient for studies with the exact
diagonalization method. Ground state properties of the planar model on
a square lattice have been intensively investigated \cite{dago}, more
recently also the $T>0$ properties using the finite-temperature
Lanczos method \cite{jprev}.

Let us go straight to results at the intermediate (optimum) doping,
where we can reach lowest $T_{fs}$.  Instead of $\sigma(\omega)$
directly it is more instructive to present the current correlation
function $C(\omega)$, equation (\ref{reg}). To avoid the ambiguities
with an additional smoothing, we plot the corresponding integrated
spectra
\begin{equation}
I_C(\omega)=\int_0^{\omega}
C(\omega') d\omega'. \label{eo1}
\end{equation} 
In Fig.~6 we present $I_C(\omega)$ for the intermediate (optimal)
doping $c_h=3/16$, for various $T\le t$.  Spectra reveal several
remarkable features \cite{jprev}:

a) For $T \le J$ spectra $I_C(\omega)$ are essentially independent of
$T$, at least for available $T>T_{fs}$.

b) Simultaneously the slope of $I_C(\omega < 2~t)$ is nearly constant,
i.e.  we find $C(\omega) \sim C_0$ in a wide range $\omega<2t $. At
the same time $C_0$ is only weakly dependent on $J$ as tested for
$J/t=0.2 - 0.6$.

c) Even for higher $T>J$ the differences in the slope $C_0$, as also
in $I_C(\infty)$, appear as less essential. 

\begin{figure}
\centering
\epsfig{file=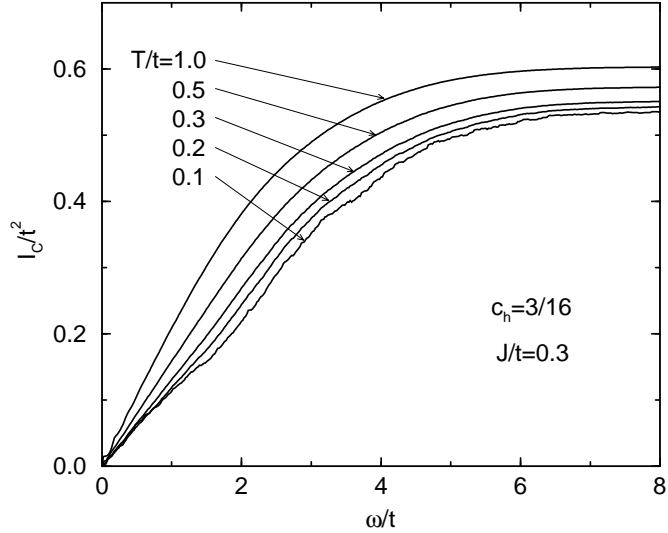,height=10cm,angle=-90}
\caption{
Integrated current correlation spectra $I_C(\omega)$ at $c_h=3/16$ for
different $T\le t$. } \label{5.4}
\end{figure}

We conclude that $C(\omega<2~t) \sim C_0$ implies a simple universal
form,
\begin{equation}
\sigma(\omega)=C_0 e_0^2\frac{1-e^{-\beta\omega}}{\omega}. \label{eo2}
\end{equation}
Such a $\sigma(\omega)$ shows a nonanalytic behaviour at $\omega
\rightarrow 0$, starting with a finite slope.  This is already an
indication that $\sigma(\omega)$ is not consistent with the usual
Drude form, but rather with a marginal concept \cite{varm} where the
only $\omega$ scale is given by $T$.  It is also remarkable that the
form (\ref{eo2}) trivially reproduces the linear law $\rho \propto T$
as well as the non-Drude fall-off at $\omega >T$. It is evident that
the expression (\ref{eo2}) is universal containing the only parameter
$C_0$ as a prefactor.

Experimental results and theoretical considerations are often
discussed in terms of the $\omega$-dependent relaxation time $\tau$
and the effective mass $m^*$. These can be uniquely introduced via the
complex $\tilde \sigma(\omega)$ and the corresponding memory function
$M(\omega)$,
\begin{equation}
\tilde \sigma(\omega)= \frac{ie_0^2 {\cal S} }{\omega + M(\omega)}, \qquad
\qquad {\cal S} = -\langle H_{kin} \rangle/2N, \label{eo3}
\end{equation}
and
\begin{eqnarray}
\frac{1}{\tau(\omega)} &=&\frac { M''(\omega)}{1+ M'(\omega)/ \omega},
\nonumber \\
\frac{m^*(\omega) }{m_t} &=& \frac {2 c_h t} {{\cal S}}
\left(1+\frac{M'(\omega)}{\omega}\right), \label{eo4}
\end{eqnarray}
where $m_t = 1/2ta_0^2$ is the bare band mass. Using
relations (\ref{eo3}),(\ref{eo4}) one can formally rewrite $\tilde
\sigma(\omega)$ in the familiar Drude form
\begin{equation}
\tilde \sigma(\omega) = \frac{i c_h e_0^2 }{m^*(\omega)[\omega +i
/\tau(\omega)]}. \label{eo5}
\end{equation}
Employing the equation (\ref{eo5}) we evaluate both $\tau(\omega)$ and
$m^*(\omega)$ from known $\tilde \sigma(\omega)$.  It follow directly
from the form (\ref{eo2}), that in the regime $T < J$ and $\omega<t$
we can describe the behaviour of $1/\tau$ with a linear law
\begin{equation}
\tau^{-1} = 2\pi\lambda (\omega + \xi T), \qquad \lambda
\sim 0.09,~~~\xi \sim 2.7~. \label{eo6}
\end{equation}
This dependence falls within the general framework of the marginal
Fermi liquid scenario \cite{varm}.  It should be however stressed that
the asymptotic form (\ref{eo6}) does not allow for any free parameter,
i.e. constants $\lambda$ and $\xi$ are universal and independent of
any model parameters, whereas within the original marginal
Fermi-liquid proposal $\lambda$ is an adjustable parameter while $\xi
= \pi$.

\begin{figure}[ht]
\centering
\epsfig{file=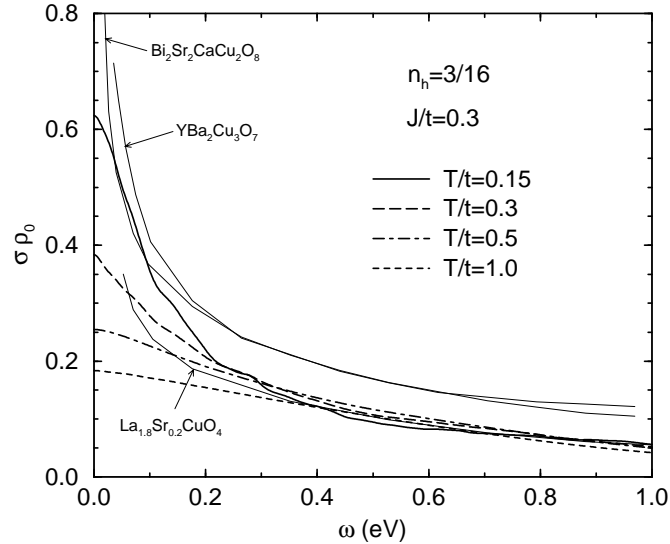,height=10cm,angle=-90}
\caption{
Sheet conductivities $\sigma(\omega)$ for various $T/t$, in comparison
with measurements in different cuprates. Experimental results refer to
$T<200~K$.  }
\end{figure}

Let us relate obtained model results to experimetns in cuprates.
Since parameters of the model, corresponding to cuprates, are to large
extent known, i.e. $J/t \sim 0.3$ and $t \sim 0.4$eV, the direct
comparison of $\sigma(\omega$ and $\rho(T)$ can be performed without
any additional assumptions. In 2D $\sigma$ is naturally expressed in
terms of the universal constant $\rho_0=\hbar/e_0^2$.  The
corresponding 3D conductivity of a stack of 2D conducting sheets with
an average distance $\bar{c}$ is given instead by $\sigma_{3D}=
\sigma/\bar c$.  For comparison with experiments we reduce the 3D
measured values to the 2D conductivities for three different cuprates
at intermediate doping, i.e. La$_{2-x}$Sr$_x$CuO$_4$ (LSCO) with
$x=0.2$, Bi$_2$sr$_2$CaCu$_2$O$_{8+\delta}$ (BSCCO) and
YBa$_2$Cu$_3$O$_7$ (YBCO). Taking into account the uncertainty in
effective hole doping within these materials calculated spectra
$\sigma(\omega)$ for $c_h=3/16$ are in a quantitative agreement with
measurements, as seen in Fig.~7.It should be noted that also
calculated $\tau^{-1}(\omega)$ and corresponding parameters $\lambda$
and $\xi$, as defined by the equation (\ref{eo6}) are close to the
experimental ones.

\begin{figure}
\centering
\epsfig{file=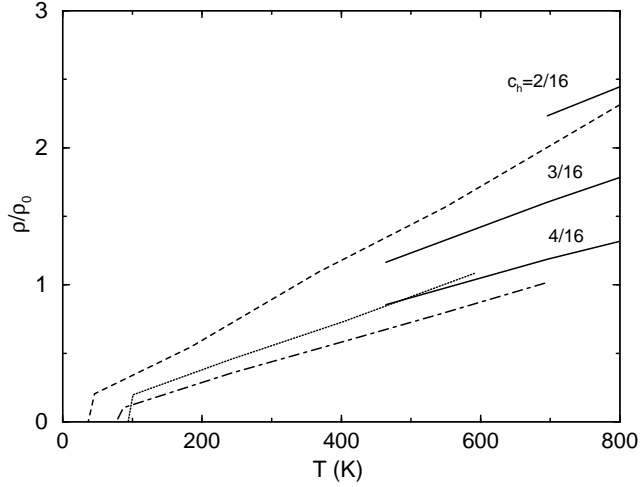,height=10cm,angle=-90}
\caption{
Sheet resistivities $\rho(T)$ for various dopings (full lines) in
comparison with measurements in LSCO with $x=0.15$ (dotted), BSCCO
(dashed), and YBCO (dash-dotted).  } \label{5.10}
\end{figure}

In Fig.~8 we compare calculated $\rho(T)$ to the measured
ones. It should be pointed out that there is a restricted $T$ window
where a comparison could be made since $T_{fs} \sim 450 K$, whereby at
$T\sim T_{fs}$ also finite-size effects start to influence our
analysis.  Nevertheless, for the intermediate doping $c_h \sim 0.2$
our $\rho(T)$ results match quantitatively well experimental ones for
cuprates with comparable hole concentrations, i.e. for BSCCO, YBCO
and LSCO with $x=0.15$.

\subsection{Origin of anomalous conductivity} 

There is so far no accepted explanation for the origin of the
anomalous marginal-type charge dynamics in cuprates. From our analysis
of the $t$-$J$ model at the intermediate doping it seems to be quite
generic feature of frustrated regime of correlated electrons. In the
following we try to argue that there is in general a relation between
the marginal-type $\sigma(\omega)$ and the overdamped character of
quasiparticle excitations, also observed experimentally in cuprates via
the angle-resolved photoemission spectroscopy (ARPES) as well in model
studies.

Let us perform for $C(\omega)$ the decoupling in terms of
single-particle (electron) spectral functions $A({\bf k}, \omega)$
neglecting possible vertex corrections, i.e.
\begin{equation}
C(\omega)=\frac{2\pi e_0^2}{N} \sum_{\bf k} (v_{\bf k}^\alpha)^2 \int
{d\omega^\prime} f(-\omega^\prime) f(\omega^\prime-\omega)  
A({\bf k},\omega^\prime) A({\bf k},\omega^\prime-\omega),
\label{eq4}
\end{equation}
where $f(\omega)$ are Fermi functions and $v_{\bf k}^\alpha$ are
(unrenormalized) band-velocity components. While such an approach is
usual in weak scattering problems, qualitative features could be
reasonable also for strongly correlated electrons.  We assume also
that $A({\bf k}, \omega)$ for quasiparticles close to the Fermi energy
($\omega=0$) can be generally presented as
\begin{equation}
A({\bf k}, \omega)=\frac{1}{\pi}\frac{Z_{\bf k}\Gamma_{\bf k}}
{(\omega-\epsilon_{\bf k})^2+\Gamma^2_{\bf k}},
\label{eq55}
\end{equation}
where quasiparticle parameters $Z_{\bf k},\Gamma_{\bf k},\epsilon_{\bf
k}$ can still be dependent on $\omega$ and $T$. In order to reproduce
the marginal Fermi liquid form, Eqs.(\ref{eo5}),(\ref{eo6}), of
$\sigma(\omega)$ we have to postulate an analogous form also for the
quasiparticle damping, i.e. $\Gamma = \gamma (|\omega| + \xi T)$
(consistent with recent ARPES experiments in BSCCO), but we neglect
the ${\bf k}$ dependence of $\Gamma$ and $Z$.

For $\omega<\omega^*$, whereby the cutoff $\omega^*$ appears due to
the effective bandwidth or some other characteristic quasiparticle
scale, only the behavior near the Fermi surface should be
important. Replacing in Eq.(\ref{eq4}) the ${\bf k}$ summation with an
integral over $\epsilon$ with a slowly varying density of states
${\cal N}(\epsilon)$ one can derive
\begin{equation}
\int d\epsilon A(\epsilon,\omega') A(\epsilon,\omega'-\omega)=
\frac{Z^2}{\pi} \frac {\bar \Gamma(\omega,\omega')}
{\omega^2+ \bar \Gamma(\omega,\omega')^2} , \label{eq66}
\end{equation}
where $\bar \Gamma(\omega,\omega')=\Gamma(\omega')
+\Gamma(\omega'-\omega)$.  We are thus dealing with
a function $C(\omega)$,
\begin{equation}
C(\omega)=\bar C \int d\omega' f(-\omega')f(\omega'-\omega)
\frac {\bar \Gamma(\omega,\omega')}
{\omega^2+ \bar \Gamma(\omega,\omega')^2},
\label{eq77}
\end{equation}
depending only on the ratio $\omega/T$, and on the parameters
$\gamma,\xi$. In the theory of Fermi liquids $\omega$ and $T$
variation are tightly bound so $\xi$ should also be quite universal or
at least quite restricted in range, whereby $\xi\sim \pi$ is usually
used \cite{varm}.

It is evident from Eq.(\ref{eq77}) that for $\gamma \ll 1$
we recover $C(\omega)$ strongly peaked at $\omega = 0$ and
consequently $\sigma(\omega) \propto \bar \Gamma/(\omega^2+ \bar
\Gamma^2)$ with $\bar \Gamma= 2\Gamma(\omega/2,T)$. This is just a
generalized Drude form with $1/\tau(\omega)=2\Gamma(\omega/2)$.
$\sigma(\omega)$ is in this case directly determined by the
single-particle relaxation $\Gamma$.

\begin{figure}[h]
\centering
\epsfig{file=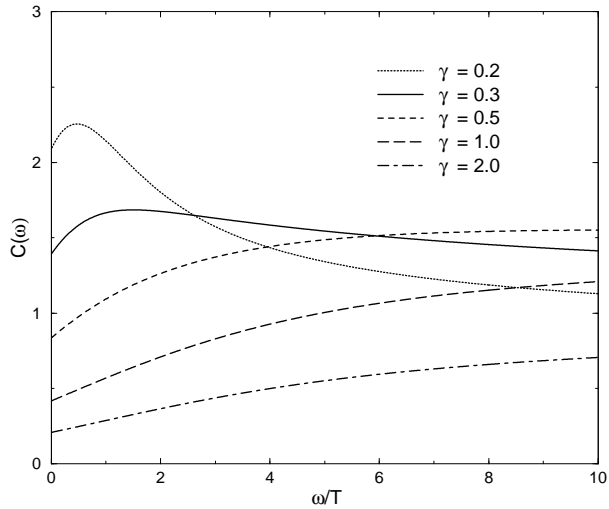,height=8cm,angle=-90}
\caption{ Current-current correlation spectra $C(\omega)$
vs. $\omega/T$ for various $\gamma$ at fixed $\xi= \pi$.}
\label{fig2}
\end{figure}

This relation is however not valid when one approaches the regime of
overdamped QP excitations $\gamma \sim 1$ or more appropriate
$\gamma\xi \sim 1$. We present in Fig.~9 results for $C(\omega)$ for
several $\gamma$ fixing $\xi=\pi$. While for $\gamma < 0.2$ still a
pronounced peak shows up at $\omega \sim 0$, $C(\omega)$ becomes
for $\gamma > 0.3$ nearly constant  or very slowly varying in a wide
range of $\omega/T$. For large $\gamma$ we find even $C(0)<C(\omega \to
\infty)$, approaching for $\gamma \gg 1$ the ratio
$C(0)/C(\infty)=1/\xi$.

The main message of the above simple analysis is that for systems with
overdamped quasiparticle excitations of the marginal Fermi-liquid form
the expression Eq.(\ref{eo2}) describes quite well $\sigma(\omega)$
for a wide range of parameters. It should be stressed that nearly
constant $C(\omega<\omega^*)$ also means that the current relaxation
rate $1/\tau^*$ is very fast, $1/\tau^* \sim \omega^* \gg 1/\tau$,
i.e. much faster than the conductivity relaxation scale apparent from
Eq.(\ref{eo6}) where $1/\tau \propto T$ is determined solely by
thermodynamics.

\section{Conclusions}

The theory of electron transport in models and materials dominated by
strong electron-electron repulsion is still in its infancy. Present
lectures try to describe some of novel phenomena in transport, which
seem to be particular to correlated systems. At the same time it is
evident that there are even more open problems which wait for proper
theoretical understanding and feasible methods for systematic study:

a) The singular transport in nontrivial integrable models of
interacting fermions reveal an evident question which are relevant
scattering processes which determine transport quantities. Coulomb
repulsion and Umklapp processes are in general clearly not enough to
lead to current relaxation and dissipation. As shown in integrable
systems transport quantities remain singular. It is expected that in
general any deviation from integrability will lead to a normal
resistence, however there exists no reliable approach to transport in
the vicinity of integrable points.

b) Even in integrable 1D systems there are several open questions.
Although the relation to conserved quantities is very appealing and
more operational it is questionable whether it can explain generally
the existence of ideal conductors, e.g. the ideal conductor in the
simplest cases as the $t$-$V$ model at half filling, the particle in
the fermionic bath etc.

c) The existence of very unusual ideal insulator has not been proven
nor disproven yet.

d) For higher dimensional systems there is definite need for
analytical methods which would allow the evaluation of (at least
qualitatively correct) transport quantities. Recently developed
numerical methods for finite correlated systems, in particular the
finite-temperature Lanczos method, have proven to be very
useful. Their weakness is clearly in the smallness of reachable
systems.  Fortunately, the physics in several relevant models appears
to be quite local in character so that e.g. mean free path is very
short even at moderately low temperatures.

e) The main motivation to study correlated systems comes from the
observation of anomalous conductivity $\sigma(\omega)$ in high-$T_c$
cuprates as prominent example of materials with strongly correlated
electrons. Numerical studies as well as analytical relation with
spectral functions with overdamped quasiparticles confirm the observed
behavior descirbed within the marginal Fermi liquid scenario. These
materials seem to follow a novel quantum diffusion law where
relaxation is dominated by thermodynamics only. However its origin is
not well understood. Also it is not clear how general such scenario
might be and whether (why) it breaks down at low enough temperatures.

\end{document}